\documentclass{aa}
\usepackage{graphicx}
\normalfont

\def \A{{\rm A}}
\def\etal{{\it et~al. }}
\def\J2{$J_{2\odot}$}
\def\cmr2{$C/MR^{2}$}
\def\32{$3$:$2$}

\begin{document}

     \title{Theory of the Mercury's spin-orbit motion and analysis 
     of its main librations}
     \author{N. Rambaux \and E. Bois }
     \offprints{rambaux@obs.u-bordeaux1.fr}
     \institute{Observatoire Aquitain des Sciences de l'Univers,
                    Universit\'e Bordeaux 1, UMR CNRS/INSU 5804, B.P. 89, F-33270\\
                    Floirac, France}
     \date{Received / Accepted }

 \abstract{      
 The \32 spin-orbit resonance between the rotational and orbital motions of 
 Mercury (the periods are $P_{\phi}$ = 56.646 and $P_{\lambda}$ = 87.969
 days respectively) results from a functional dependance of the tidal friction 
 adding to a non-zero eccentricity and a permanent asymmetry in the equatorial 
 plane of the planet. The upcoming space missions, MESSENGER and 
 BepiColombo with onboard instrumentation capable of measuring the rotational
 parameters stimulate the objective to reach an accurate theory of the rotational 
 motion of Mercury. For obtaining the real motion of Mercury, we have 
 used our BJV model of solar system integration including the coupled spin-orbit
 motion of the Moon. This model, expanded in a relativistic framework, had been 
 previously built in accordance with the requirements of the Lunar Laser Ranging 
 observational accuracy. We have extended the BJV model by generalizing the 
 spin-orbit couplings to the terrestrial planets (Mercury, Venus, Earth, and Mars).
 The updated model is called SONYR (acronym of Spin-Orbit N-BodY Relativistic model). 
 As a consequence, the SONYR model gives an accurate simultaneous integration of the 
 spin-orbit motion of Mercury. It permits to analyze the different families 
 of rotational librations and identify their causes such as the planetary interactions or the 
 parameters involved in the dynamical figure of the planet. The spin-orbit motion of 
 Mercury is characterized by two proper frequencies (namely $\Phi$ = 15.847 years and $\Psi$
 = 1066 years) and its 3:2 resonance presents a second synchronism which can be understood as
  a \textit{spin-orbit secular resonance} ($\Pi$ = 278\,898 years). A new determination
 of the mean obliquity is proposed in the paper. By using the SONYR model, we
 find a mean obliquity of 1.6 arcminutes. This value is consistent with the Cassini 
 state of Mercury.  Besides, we identify in the Hermean librations the impact of 
 the uncertainty of the greatest principal moment of inertia (\cmr2) on the obliquity and 
 on the libration in longitude (2.3 milliarcseconds and 0.45 arcseconds respectively for an increase 
 of 1$\%$ on the \cmr2 value). These determinations prove to be suitable for providing 
 constraints on the internal structure of Mercury.  
 
  }
  
     \maketitle

\section{Introduction}\label{sec:introduction}

Before 1965, the rotational motion of Mercury was assumed to be
synchronous with its orbital motion. In 1965, Pettengill \& Dyce discovered 
a 3:2 spin-orbit resonance state by using Earth-based radar
observations (the Mercury's rotation period is $P_{\phi}$ = $58.646$ days
while the orbital one is $P_{\lambda}$ = $87.969$ days). This surprising resonance
results from a non-zero eccentricity and a permanent asymmetry in 
the equatorial plane of the planet. In addition, the 3:2 resonance strongly 
depends on the functional dependance of the tidal torque on the rate of the 
libration in longitude. Moreover the 3:2 resonance state is preserved by the
tidal torque (Colombo \& Shapiro 1966). The main dynamical features of
Mercury have been established during the 1960's in some pioneer works such
as Colombo (1965),  Colombo \& Shapiro (1966), Goldreich \& Peale (1966) 
and Peale (1969). Goldreich \& Peale (1966) notably studied the probability of
resonance capture and showed that the 3:2 ratio is the only possible one for a 
significant probability of capture. In addition, in a tidally evolved system, the 
spin pole is expected to be trapped in a Cassini state (Colombo 1966; Peale 1969,
1973). The orbital and rotational parameters are indeed matched in such a way 
that the spin pole, the orbit pole, and the solar system invariable pole 
remain coplanar while the spin and orbital poles precess. The reader may find in
Balogh \& Campieri (2002) a review report on the present knowledge of Mercury
whose the interest is nowadays renewed by two upcoming missions: 
MESSENGER (NASA, Solomon \etal 2001) and BepiColombo (ESA, ISAS, 
Anselin \& Scoon 2001).

Our work deals with the physical and dynamical causes that contribute to
induce librations around an equilibrium state defined by a Cassini state.
In order to wholly analyze the spin-orbit motion of Mercury and
its rotational librations, we used a gravitational model of the
solar system including the Moon's spin-orbit motion. The framework of the
model has been previously constructed by Bois, Journet \& Vokrouhlick\'y 
(BJV model) in accordance with the requirements of Lunar Laser Ranging 
(LLR thereafter) observational accuracy (see for instance a review 
report by Bois 2000). The approach of the model consists in integrating the
\textit{N}-body problem on the basis of the gravitation description given by 
the Einstein's general relativity theory according to a formalism derived from
the first post-Newtonian approximation level. The model is solved by modular
numerical integration and controlled in function of the different physical 
contributions and parameters taken into account. We have extended this model
to the integration of the rotational motions of the terrestrial planets (Mercury, Venus, 
Earth, and Mars) including their spin-orbit couplings. The updated  model 
is then called SONYR (acronym of Spin-Orbit \textit{N}-bodY Relativistic 
model). As a consequence, using SONYR, the \textit{N}-body problem for
the solar system and the spin motion of Mercury are simultaneously integrated. 
Consequently we may analyze and identify the different families of Hermean rotational
librations with the choice of the contributions at our disposal.

Starting with the basic spin-orbit problem according to Goldreich \& Peale 
(1966), we have computed a surface of section for the Mercury's rotation 
showing its very regular behavior. We have calculated again the proper 
frequency for the spin-orbit resonance state of Mercury.
Using our model, an important part of the present study deals with the
main perturbations acting on the spin-orbit motion of Mercury such
as the gravitational figure of the planet as well as the planetary effects
and their hierarchy. A detailed analysis of the resulting rotational librations
due to these effects is presented and described in the paper. 
A new determination of the Hermean mean obliquity is also proposed.
Moreover, we identify in the Hermean librations the impact
of the variation of the greatest principal moment of inertia on the 
instantaneous obliquity and on the libration in longitude.
Such a signature gives noticeable constraints on the internal structure of Mercury. 

\section{Geometry of the spin-orbit coupling problem}\label{sec:analyse}

\begin{figure}[!ht]
      \begin{center}
        \hspace{0cm}
        \includegraphics[width=6cm]{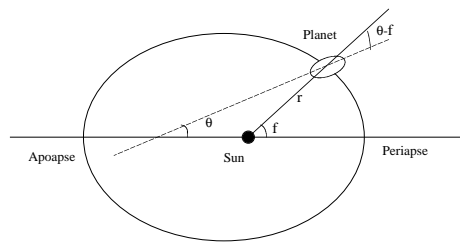}
        \caption{Geometry of the spin-orbit coupling problem. 
	$r$ is the radius vector, $f$ 
	the true anomaly and $\theta$ is the characteristic
	angle of rotation.}
        \label{geometry}
      \end{center}
\end{figure}

According to Goldreich \& Peale (1966), we consider the spin-orbit
motion of Mercury with its spin axis normal to the orbital plane.
The orbit is assumed to be fixed and unvariable (semi-major
axis $a$ and its eccentricity $e$). The position of Mercury is determined by 
its instantaneous radius $r$ while its rotational orientation is specified by the 
angle $\theta$. The orbital longitude is specified by the true anomaly 
$f$ while the angle $\theta - f$ measures the angle between
the axis of least moment of inertia of Mercury and the Sun-to-Mercury line 
(see Fig.~\ref{geometry}). According to these assumptions, the dynamical
problem of the spin-orbit motion of Mercury is reduced to a one-dimensional 
pendulum-like equation as follows~:
\begin{equation}
          C\ddot{\theta} + \frac{3}{2} (B-A) \frac{GM_{\odot}}{r^{3}}
      \sin{2(\theta-f)}  = 0
      \label{eq:spinorb}
\end{equation}
where $G$ is the gravitational constant, $M_{\odot}$ the solar mass,
and $A, B$, and $C$ the principal moments of inertia of Mercury.
This equation of motion has only a single degree of freedom for the spin-orbit
coupling, the characteristic angle of rotation $\theta$, but depends explicitly on time
through the distance $r$ to the planet and the non-uniform Keplerian motion of the true 
anomaly $f$. As a consequence, it is a problem not reducible by quadrature
and the equation (\ref{eq:spinorb}) is non-integrable (Wisdom $1987$).


\begin{figure*}[!htb]
     \begin{center}
        \hspace{0cm}
        \includegraphics[width=18cm]{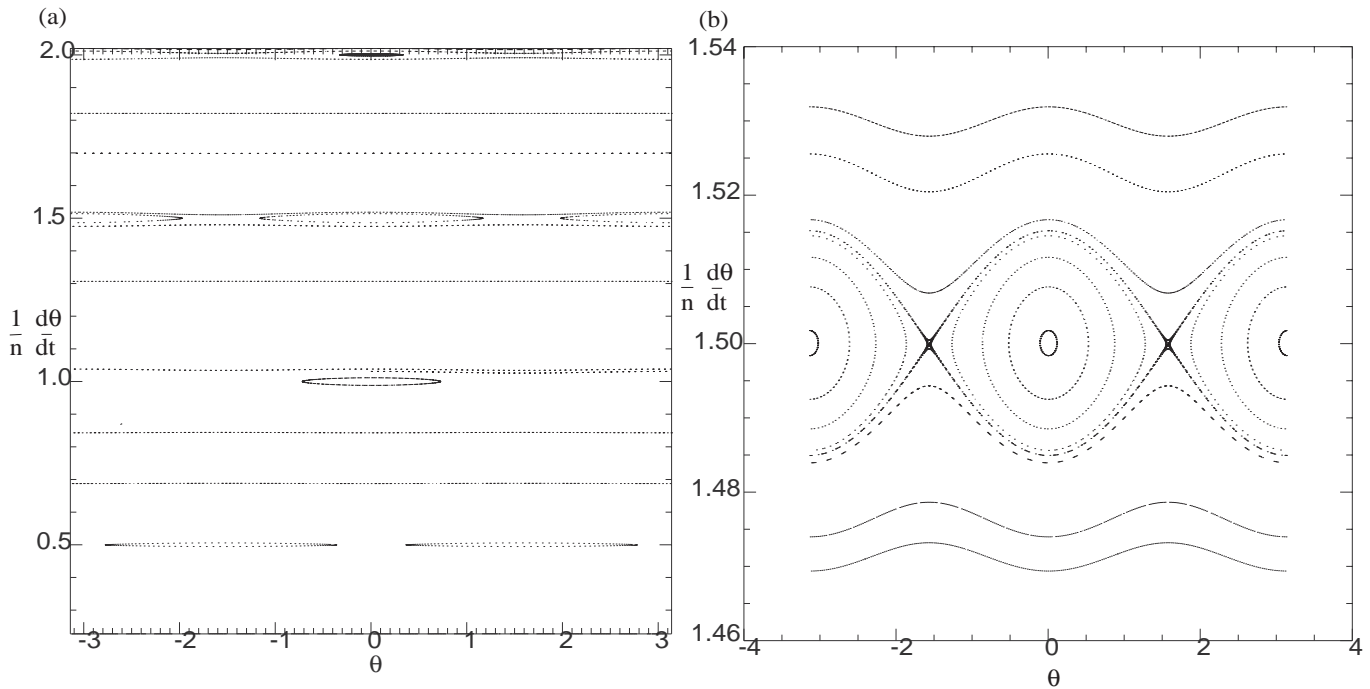}
        \caption{Surface of section of the Mercury's spin-orbit 
        coupling ($\alpha = 0.0187,
        e = 0.206$). The rate of change of the orientation is plotted versus
        the orientation of Mercury defined by the angle $\theta$ at every pericenter
        passage. The spin axis is constrained
        to be normal to the orbital plane. The plot (a) makes
        in evidence the different quasiperiodic librations
        around the 1:2, 1:1 (synchronous), 3:2, and 2:1 rotation states.
        The plot (b) is a zoom of the real 3:2 resonance state of Mercury.
        The chaotic zone is microscopic and non visible at this scale.}
        \label{mercure}
      \end{center}
\end{figure*}


In order to know the structure of the phase space of the Mercury's
rotation, a surface of section for its spin-orbit coupling is very 
useful. Let be $\alpha=\sqrt{ \frac{3(B-A)}{C} }$  the asphericity of the
Mercury's dynamical figure combining the principal moments of inertia
$A, B$, and $C$. The equation~(\ref{eq:spinorb}) becomes~:
\begin{equation}
       \ddot{\theta} +
       \frac{1}{2}n^{2}\left(\frac{a}{r}\right)^{3}\alpha^{2}\sin2(\theta -f)=0
       \label{section}
\end{equation}
where $n$ is the orbital mean motion, $f$ and $\theta$ as defined above.
The numerical integration of the equation (\ref{section}) yields to
the Figure~\ref{mercure} showing stroboscopically, i.e. once point
per orbit at the periapse, various trajectories of the phase portrait of
Mercury ($\theta$ and $1/n d \theta / dt$ on the axes $x$ and $y$
respectively). Figure~(\ref{mercure}.a) illustrates the
different quasiperiodic librations around the 1:2, 1:1 (synchronous), \32,
and 2:1 rotation states while Figure~(\ref{mercure}.b) shows a zoom
of the present \32 resonance state for Mercury. Let us notice that in the
1:2 rotation state, the two libration areas are shifted by $\pi/2$ with
respect to the other rotation states. 
The former rotation rate of Mercury was higher and the planet could
have experienced large and chaotic variations in obliquity at some time
of the past (Laskar \& Robutel 1993). Due to its slowing down by tidal
despin, the Mercury's rotation has then encountered different resonance
states, from higher orders to the one trapped in the present state. 
Goldreich \& Peale $(1966)$ have shown that this present state 
(the \32 resonance) was the first one with a substantial probability of capture.
This one indeed depends crucially on the functional dependence of the tidal
torque acting on the spin-orbit motion. This dependence is expressed by the 
variation rate of the longitude libration angle $\gamma$ as defined below. 
In the end, the stabilization of Mercury at the \32 spin-orbit resonance is due
to permanent asymmetry on the equatorial plane, as well as its non-zero
eccentricity equal to $0.206$ (Colombo 1966, Colombo \& Shapiro 1966).

According to the Chirikov resonance overlap criterion (1979), the chaotic
behavior appears when the asphericity of the body is larger than the
following critical value~:
\begin{equation}
      \alpha^{cr} =  \frac{1}{2 + \sqrt{14e}}
\end{equation}
where $e$ is the orbital eccentricity. In the case of Mercury,
$\alpha = 0.0187$ is lower than the critical value $\alpha^{cr} = 0.2701$.
As a consequence, the spin-orbit behavior is regular.
The zoom in Figure (\ref{mercure}. b) shows indeed that the separatrix
surrounding the \32 resonance state is very small. The width of the
associated chaotic zone is then estimated to the order of $10^{-43}$
(Wisdom \etal 1984). Consequently, tidal friction pulls Mercury accross
the chaotic separatrix in a single libration period.

From the equation (\ref{eq:spinorb}), it is possible to obtain an integrable
approximated equation using the spin-orbit resonance, the spin rate 
$\dot{\theta}$ being commensurable with the mean orbital motion $n$. Following 
Murray \& Dermott ($2000$), by introducing a new variable $\gamma= 
\theta -pn$ where $p$ parametrizes the resonance ratio ($p = \frac{3}{2}$
in the case of Mercury), one may expand the equation in form-like Poisson 
series. Taking into account that $\dot\gamma \ll n $, one averages all the
 terms over one orbital period, and finally obtain the
following equation~:
\begin{equation}
          \ddot{\gamma} + \frac{3}{2} n^{2}\frac{B-A}{C}H(p,e)
      \sin{2\gamma}  = 0
      \label{bg}
\end{equation}
where $H(p,e)$ is a power series in eccentricity. In the case of Mercury,
this expression is written as follows~:
\begin{equation}
      H(\frac{3}{2},e) =  \frac{7}{2}e - \frac{123}{16} e^{3} \\
\end{equation}
Finally, the proper frequency of the Mercury's spin-orbit motion is~:
   \begin{equation}
       \omega_{0}  =   n \left[ 3 \frac{B-A}{C} | H(p,e) |
       \right]^{\frac{1}{2}} \\
   \end{equation}
which by using the values listed in Table~\ref{paratot} gives 
the proper period of $15.830$ years.

\begin{figure}[!htb]
      \begin{center}
        \hspace{0cm}
        \includegraphics[width=6cm,angle=-90]{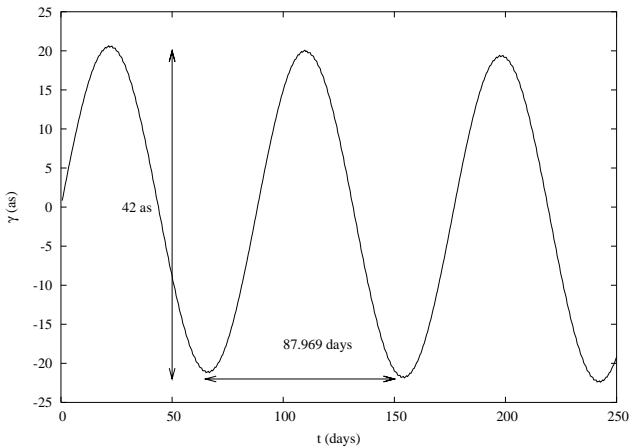}
        \caption{The spin-orbit solution of Mercury in the planar case
        (equation~\ref{bg2}) plotted over 250 days. Arcseconds are 
        on the vertical axis and days on the horizontal axis. Short-term 
        librations have a period of 87.969 days (the orbital period of 
        Mercury) and 42 as of amplitude.}
        \label{plbg}
      \end{center}
\end{figure}

Balogh \& Giamperi (2002) developed the equation (\ref{section}) and obtained 
 the following expression~:
\begin{equation}
          \ddot{\gamma} + \alpha_{0} \sum_{q}G_{20q}(e) \sin[2 \gamma + 
          (1-q)M] = 0
      \label{bg2}
\end{equation}
where the $G_{20q}$ coefficients are eccentricity functions defined
by Kaula (1966). The $G_{201}$ coefficient is equal to
$H(\frac{3}{2},e)$. $\alpha_{0} = \frac{3}{2} \frac{B-A}{C}$ = 
1.76 $10^{-4}$ for the Mercury case. Figure~\ref{plbg} 
presents then the numerical solution of the formula (\ref{bg2}) (thanks to 
G. Giampieri,  private communication). The angle $\gamma$ describes
a periodic behavior with a period equal to the revolution of Mercury. 
The amplitude of 42 arcseconds (as) depends on the value of $\alpha_{0}$.
The behavior of the $\gamma$ angle is nicely
matched by the approximate formula of Jhen \& Corral (2003)~:
\begin{equation}
    \gamma = \phi_{0} \sin{M} + \phi_{1} \sin{2M} 
    \label{eq:rudi}
\end{equation}
where $\phi_{0}=\alpha_{0} (G_{200}-G_{202})$ and $\phi_{1}=  
\frac{\alpha_{0}}{4} (G_{20-1}-G_{203})$. The authors noticed that the 
ratio of the two amplitudes ($\phi_{0}$ and $\phi_{1}$) does not depend 
on the $\alpha_{0}$ parameter, and $K= \frac{\phi_{0}}{\phi_{1}}=-9.483$.

However, the above equations describe a spin-orbit motion of Mercury 
where the spin axis is normal to the orbital plane while the orbital 
motion is Keplerian.

\section{The extended BJV model~: SONYR}\label{sec:methode}

\subsection{Theory}

In order to wholly analyze the spin-orbit motion of Mercury and
its rotational librations, we have enlarged a gravitational model
(called BJV) of the solar system including the Moon's spin-orbit
motion.
The accurate theory of the Moon's spin-orbit motion, related to
this BJV model, was constructed by Bois, Journet \& Vokrouhlick\'y
in accordance with the high accuracy of the LLR observations
(see previous papers~: Bois \etal 1992; Bois \& Journet 1993; Bois \& 
Vokrouhlick\'y 1995; Bois \etal 1996; Bois \& Girard 1999). The approach
of the BJV model consists in integrating the \textit{N}-body problem 
(including translational and rotational motions) on the basis of the gravitation
description given by the Einstein's general relativity theory.
The equations have been developped in the DSX formalism
presented in a series of papers by Damour, Soffel \& Xu (Damour \etal
1991, 1992, 1993, 1994). For purposes of celestial mechanics,
to our knowledge, it is the most suitable formulation of the post-Newtonian
(PN) theory of motion for a system of $N$ arbitrary extended, weakly 
self-graviting, rotating and deformable bodies in mutual interactions. 
The DSX formalism, derived from the first post-Newtonian approximation
level, gives the post-Newtonian representation of the translational motions of
the bodies as well as their rotational ones with respect to the locally transported 
frames with the bodies.

Gravitational fields of the extended bodies are parameterized in
multipole moment expansions~: $(M^\A_L, S^\A_L)$ define the mass and spin
Blanchet-Damour multipoles characterizing the PN gravitational field of
the extended bodies while $(G^\A_L, H^\A_L)$ are tidal gravitoelectric
and gravitomagnetic PN fields. Because we do not dispose of dynamical
equations for the quadrupole moments $M^\A_{ab}$, and although the
notion of rigidity faces conceptual problems in the theory of relativity,
we have adopted the `rigid-multipole' model of the extended bodies as
known from the Newtonian approach. Practically this is acceptable since
the relativistic quadrupole contributions are very small. Consequently
and because it is conventional in geodynamical research to use spherical
harmonics analysis of the gravitational fields with the corresponding
notion of harmonic coefficients $(C^\A_{lm}, S^\A_{lm})$, the quadrupole
moments $M^\A_{ab}$ have been expressed in those terms, according to
reasons and assumptions given in Bois \& Vokrouhlick\'y (1995).
Gravitational figures as well as the figure-figure interactions of
the bodies are then represented by expansions in spherical harmonics
(Borderies 1978, Shutz 1981). Moreover, internal structures of solid
deformable bodies, homogeneous or with core-mantle interfaces, are
represented by several terms and parameters arising from tidal
deformations of the bodies (both elastic and anelastic). More details
and references on these topics are given in the above quoted papers
related to our works concerning the theory of the Moon's spin-orbit
motion.

The BJV model, as described above, has been extended to the spin-orbit
integration of the terrestrial planets (Mercury, Venus, Earth, and Mars). 
This new model is henceforth called SONYR (for Spin-Orbit 
\textit{N}-bodY Relativistic model). In the present paper framework,
the SONYR model is devoted to the detailed analysis of the complete
spin-orbit motion of Mercury.

The simultaneous integration of the solar system, including the
Mercury's spin-orbit motion, uses a global reference system given
by the solar system barycenter. Nevertheless, let us recall that local
dynamically non-rotating frames show a slow (de Sitter) rotation with
respect to the kinematically non-rotating frames. As a consequence,
the reference frame for the Mercury's rotation is affected by a slow
precession of its axes transported with the translational motion of
Mercury. In the Earth's case, the de Sitter secular precession of the
Earth reference frame is close to 1.92 as/cy (see Fukushima
1991, Bizouard \etal 1992, Bois \& Vokrouhlick\'y 1995).
Consequently, the real rotation of Mercury has not to be expressed
in an inertial system fixed in space, but in a local dynamically
non-rotating frame fallen down in the gravitational field of the Sun.
Because of the proximity of Mercury to the Sun, its de Sitter
precession may be expected quite significant.

In the end, the SONYR model and its analysis method take
into account (i) the experience in post-Newtonian gravitation in the
definition of reference frames required to deal with rotational motions
combined with translational ones, and (ii) the modern knowledge of
dynamical systems for studiing librations as quasi-periodic solutions
according to the axiomatic presented in Bois (1995).
We can state that the model is not Newtonian but rather
`Newtonian-like', resulting from truncation of the fully post-Newtonian
(DSX) framework. In the present paper, we deal with the Newtonian-like
librations (classical physical librations), while the formally
relativistic contributions (relativistic librations and de Sitter
precession of the Mercury's reference frame) will be analyzed in a
forthcoming paper.

\subsection{Method}

The model is solved by modular numerical integration and controlled in
function of the different physical contributions and parameters taken into
account. The \textit{N}-body problem (for the translational motions),
the rotational motions, the figure-figure and tidal interactions between
the required bodies are simultaneously integrated with the choice of the
contributions and truncations at our disposal.
For instance, the upper limits of the extended figure expansions
and mutual interactions may be chosen as follows~: up to $l=5$ in the Moon
case, $4$ for the Earth, $2$ for the Sun while only the Earth-Moon
quadrupole-octupole interaction is taken into account (see previous
papers). The model has been especially built to favor a systematic analysis
of all the effects and contributions. In particular, it permits the separation
of various families of librations in the rotational motions of the bodies.

The non-linearity features of the differential equations, the degree of
correlation of the studied effect with respect to its neighbors (in the
Fourier space) and the spin-orbit resonances (in the Moon and 
Mercury's cases), make it hardly possible to speak about `pure' effects with
their proper behavior (even after fitting of the initial conditions).
The effects are not absolutely de-correlated but relatively isolated.
However, the used technique (modular and controlled numerical integration,
differentiation method, mean least-squares and frequency analysis) gives the 
right qualitative behavior of an effect and a good quantification for this
effect relative to its neighbors. In the case of the particular status of the
purely relativistic effetcs, their quantitative behaviors are beyond the scope
of the present paper and will be discussed in a forthcoming work.
When a rotational effect is simply periodic, a fit of the initial conditions
for a set of given parameters only refines without really changing the
effect's behavior. The amplitudes of librations plotted in 
Figures~\ref{obl} and~\ref{lib} are then slightly upper bounds.

The precision of the model is related to the one required by the theory of the
Moon. One of the aims in building the BJV model (at present included inside
SONYR) was to take into account all phenomena up to the precision level resulting from
the LLR data (i.e. at least $1$ cm for the Earth-Moon distance, $1$ 
milliarcsecond (mas) for the librations). For reasons of consistency, several
phenomena capable of producing effects of at least $0.1$ mas had been also
modeled (the resulting libration may be at the observational accuracy level). 
Moreover, in order to justify consistence of the Moon's theory, this one had 
been adjusted to the JPL ephemeris on the first $1.5$ years up to a level of a 
few centimeter residuals. In the other hand, the internal precision of the model 
is only limited by the numerical accuracy of the integration. Thus, in order to 
avoid numerical divergence at the level of our tests for Mercury, computations 
have been performed in quadrupole precision ($32$ significant figures, integration
at a $10^{-14}$ internal tolerance).

\begin{table*}[!htb]
      \centering
       \caption{Our initial conditions at 07.01.1969 (equinox 
       J2000). (a) Mean values derived from the SONYR 
       model;
       (b) Seidelmann \etal (2002)}
      \begin{tabular}{rclrclrcl}
         \hline    
           \hline      
           \multicolumn{9}{c}{Mercury } \\
       \multicolumn{9}{c}{ Rotation angles}\\
           \hline
          & & & & & & & \\	   
       $ \psi     $ &  = &   $48.386$~deg    & 
       $\dot \psi$ & = & 0.0~deg/day &
       $< \dot \psi >$ & = & -0.616 10$^{-7}$~deg/day (a) \\
       $ \theta  $& =   & $ 7.005$~deg    &
       $\dot \theta$ &= &  0.0~deg/day & 
       $< \dot \theta >$ &= &   -0.267 10$^{-8}$~deg/day (a)       \\
       $\varphi $& = &$299.070$~deg    & 
       $\dot \varphi$& =& 6.138505~deg/day (b) & 
       $\eta_{0}$  & =&  1.6 amin (a)  \\
              \hline
       \end{tabular}
    \label{paraci}
\end{table*}

\begin{table*}[!htb]
      \centering
        \caption{Parameters of  Mercury. (a) JPL; (b) Anderson \etal 
      (1987); (c) Milani \etal 2001.}
      \begin{tabular}{lcr}
       \hline
          \hline
          \multicolumn{3}{c}{Mercury}\\
       \multicolumn{3}{c}{Bulk quantities }\\
       \hline
        & &\\
        Mass ($GM_{\odot}$) & = &  4.9125.10$^{-10}$ ${(a)}$\\
        Equatorial radius (km)&= & 2439 ${(b)}$ \\
        $J_{2}$      & = & $ (6.0 \pm 2.0) .10^{-5}$ ${(b)}$ \\ 
        $C_{22}$   & = & $ (1.0 \pm 0.5) .10^{-5}$ ${(b)}$ \\
        \cmr2         & = & 0.34 ${(c)}$  \\
        & & \\
               \hline
        \end{tabular}
    \label{paratot}
\end{table*}

\begin{table*}[!htb]
      \centering     
      \caption{Our results for the spin-orbit motion of Mercury. The 
      spin-orbit period verifies the relation~: $\widetilde{P} = 2 P_{\lambda}=3 P_{\varphi}$.}
      \begin{tabular}{lcr}
             \hline  
             \hline	
     \multicolumn{3}{c}{Mercury}\\
       \multicolumn{3}{c}{Spin-orbit characteristic period}\\
       \hline
       & &\\
        $\Phi$ (1$^{st}$ proper frequency) & = & 15.847 years \\
        $\Psi$ (2$^{nd}$ proper frequency) & = & 1066 years \\
        $\Pi$ (orbital precession) & = & 278 898 years \\
        $P_{\lambda}$ (orbital period) &  = & 87.969 days \\
        $P_{\varphi}$ (rotational period) & =  & 58.646 days \\
        $\widetilde{P}$ (spin-orbit period) & = & 176.1 days \\
       \hline        
        \end{tabular}

    \label{periode}
\end{table*}

\subsection{Terminology}

In order to de-correlate the different librations of Mercury, we use
the terminology proposed in Bois (1995), which is suitable for a general and
comparative classification of rotational motions of the celestial
solid bodies. This terminology derives from a necessary re-arrangement of
the lunar libration families due to both progress in the Moon's motion
observations (LLR) and modern knowledge of dynamical systems.

Traditionally, the libration mode called physical libration is split up
according to the conventional dualism {\it "forced-free"}.
The {\it forced} physical librations are generally related to gravitational
causes while the {\it free} librations would be departures of the
angular position from an equilibrium state.
These cuttings out contain ambiguities and redundancies discussed in
previous papers (Bois 1995, Bois 2000).
Formally, the free librations are periodic solutions of a dynamical system
artificially integrable (by a convention of writing related to
specific rates of the spin-orbit resonance, for instance 1:1),
whereas the forced librations express, in space phase, quasi-periodic
solutions around a fixed point (the system is no longer integrable).
Moreover, any stable perturbed rotation of celestial solid body contains
imbricate librations of different nature, and those are too strongly
overlapped to keep the traditional classification.

In the present terminology, the libration nature, its cause and its
designation are linked up. Two great libration families serve to
define the physical librations, namely the {\it potential} librations
and the {\it kinetical} librations. They simply correspond
to a variation energy, potential or kinetical respectively.
For libration sub-classes, the designation method is extensive to any
identified mechanism (see more details in Bois 1995).
The terminology permits easily the separation of various families
(see the Moon's case described in a set of previous papers).
These librations are called {\it direct} when they are produced
by torques acting on the body's rotation. They are called {\it indirect}
when they are produced by perturbations acting on the orbital motion
of the body. Indirect librations derive from spin-orbit couplings.

A specificity of the SONYR model with its method of analysis is to 
isolate the signature of a given perturbation. The SONYR model 
allows indeed the identification of relationships between causes and 
effects including interactions between physics and dynamics, such as 
the dynamical signature of a core-mantle interaction (called {\it centrifugal 
librations}). 

\begin{figure*}[!htb]
      \begin{center}
        \hspace{0cm}
        \includegraphics[width=10cm,angle=-90]{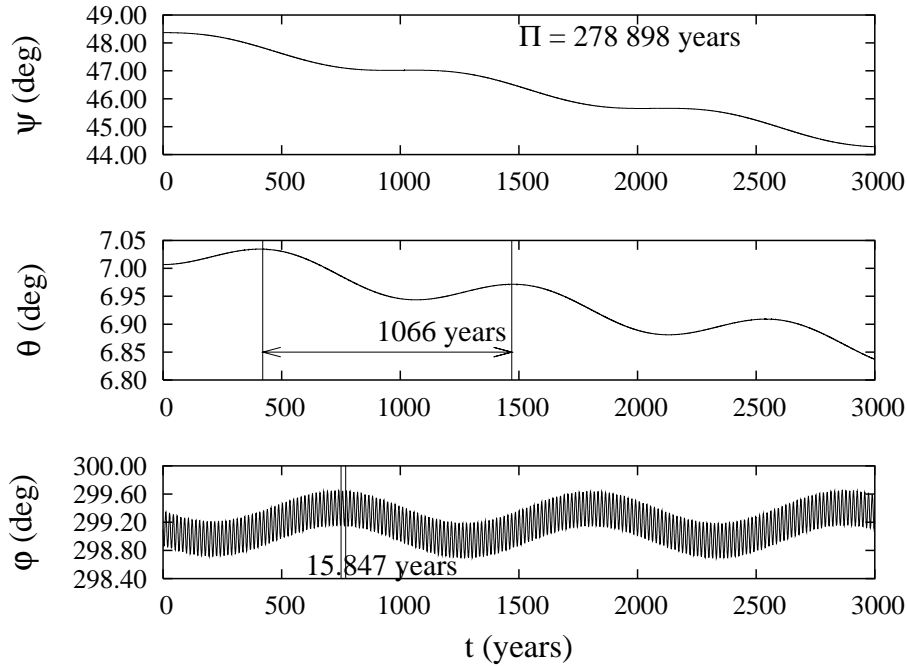}
	\vspace{-1.0cm}
        \caption{The rotational motion of Mercury expressed in the  
        ecliptic reference frame $OXYZ$ by using the 3-1-3 Eulerian 
        sequence $(\psi,\theta,\varphi)$, and plotted over 3000 years. 
        Degrees are on the vertical axis and years on the horizontal axis.
        The value of 278 898 years for the secular part of the $\psi$ angle is 
        related to the planetary interactions. The period of 1066 years 
        in the three angles is the libration of the spin-orbit 
        secular variable. The period of 15.847 years in the 
        $\varphi$ angle is the proper frequency in longitude of Mercury.}
        \label{SystSol}
     \end{center}
\end{figure*}

\subsection{Parameters and initial conditions}\label{paci}

In the computations presented in the paper, the required dynamical
parameters and general initial conditions come from the JPL DE405
ephemeris (Standish 1998). However, concerning the parameters related 
to the Mercury's rotation (second-degree spherical harmonics $C_{20}$
and $C_{22}$), which are not included in the JPL ephemeris,
our model uses those given by Anderson \etal (1987) (see
Table~\ref{paratot}). Besides, up to now it does not exist any ephemeris 
of the Mercury's rotation. As a consequence, to build initial 
conditions for the Hermean rotation (described by an Eulerian 
sequence of angles $\psi$, $\theta$ and $\varphi$ defined below in Section 
\textit{4.1}), we use the following principle~: Assuming the polar axis of 
Mercury normal to its orbital plane, we obtain $\psi = \Omega$ and $\theta = i$ 
where $\Omega$ and $i$ are respectively the ascending node and the inclination 
of the orbit of Mercury on the Earth equatorial plane (which is the reference 
frame used in the DE405 ephemeris). The long axis of Mercury being pointed
towards the Sun at its periapse allows to fix the $\varphi$ angle of 
polar rotation. The value of $\dot \varphi$ is found in Seidelmann \etal (2002). 
We use at last $\dot \psi = 0$ and  $\dot \theta = 0$; these two variables reach 
to mean values generated by the complete spin-orbit problem of Mercury~:
$-0.616 \times 10^{-7}$ deg/day and $-0.267 \times 10^{-8}$ deg/day 
respectively. The numerical integrations presented in the paper start from these 
initial conditions related to the planar problem for Mercury; they are listed 
in Table~\ref{paraci}. Departure from the planar case is understood 
as the integration of physics included in SONYR.

In the other hand, for the computations carried out in this paper,
the global reference frame $O'X'Y'Z'$ is given by a reference system 
centered on the solar system barycenter, fixed on the ecliptic plane, 
and oriented at the equinox $J2000$. The rotational motion of Mercury is
evaluated from a coordinate axis system centered on the Mercury's
center of mass $Oxyz$ relative to a local dynamically
non-rotating reference frame, $OXYZ$, whose axes are initially 
co-linear to those of $O'X'Y'Z'$. In the framework of the present paper
without purely relativistic contributions, let us note that axes of 
$OXYZ$ remain parallel to those of $O'X'Y'Z'$.

The {\it N}-body problem for the planets of the solar system and 
the Mercury's spin-orbit motion are simultaneously integrated in the 
SONYR model. Concerning the rotational equations written in a 
relativistic framework, the reader may refer to Bois \& Vokrouhlick\'y
(1995). In a Newtonian approach, these equations amount to the 
classical Euler-Liouville equations of the solid rotation (see e.g. 
Goldstein 1981). We follow the formalism and the axiomatic expanded in
Bois \& Journet (1993) and Bois (1995) for the definition of the different
rotational elements as well as the used terminology.
Let us simply precise that $\textbf{l}$ is the angular momentum expressed
in $Oxyz$ and is related to the instantaneous rotation vector 
${\bf \omega}$ as follows~:
\begin{equation}
       \textbf{l} = \left( I \right) {\bf \omega}
      \label{euler}
\end{equation}
where $\left( I \right) $ is the tensor of inertia for the body. 
According to classical assumptions,  $\left( I \right)$ is reduced 
to three principal moment of inertia $A, B$, and $C$.
The gravity field of Mercury is essentially unknown. The tracking 
data from the three fly-by of Mariner 10 in 1974-75 have been 
re-analyzed by Anderson \etal (1987) to give a low accuracy estimate of 
the normalized coefficients $C_{20}$ and $C_{22}$ 
(see Table~\ref{paratot}, values are expressed in the body-fixed frame of the 
principal axes of inertia). The principal moments of inertia $A$
and $B$ are then infered from $C_{20}$ and $C_{22}$ by the 
following formulae (Ferrari \etal 1980)~:
\begin{eqnarray}
    \frac{A}{MR^{2}} &  = & C_{20} - 2C_{22} + \frac{C}{MR^{2}}   \\
                 &   &  \nonumber \\ 
    \frac{B}{MR^{2}} & = & C_{20} + 2C_{22} + \frac{C}{MR^{2}}  \nonumber 
    \label{ABC}
\end{eqnarray}
Let us note that the parameter $C/MR^{2}$ is not constrained within these relations. 
We present in Section \textit{4.2} the variations of $C/MR^{2}$ by 
analyzing its signature into the librations of Mercury.

\section{The librations of Mercury}

\subsection{Planetary perturbations}\label{sec:stand}

\begin{figure}[!htb]
      \begin{center}
        \hspace{0cm}
        \includegraphics[width=6cm,angle=-90]{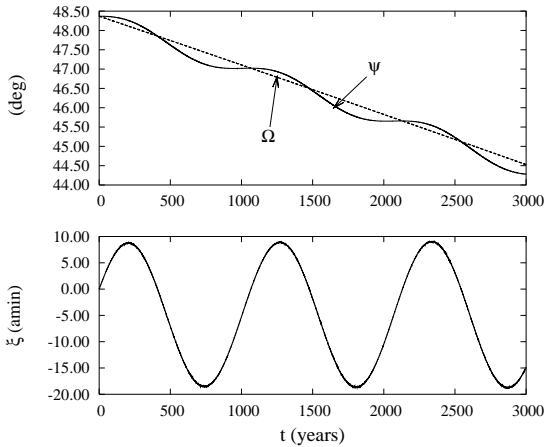}
        \caption{Mercury's spin-orbit secular resonance. The top panel presents 
        the evolution of $\psi$ and $\Omega$ both including a secular term.
        The bottom panel presents the behavior of $\xi = \psi - \Omega$ 
        plotted over 3000 years. Let us underline that this $\xi$ angle 
        does not present any secular term, showing then a synchronism
        between both angles, $\psi$ and $\Omega$.
        On the top pannel, degrees are on the vertical axis while arcminutes
        are on the bottom panel; years are on both horizontal axes.}
        \label{standardCassini}
      \end{center}
\end{figure}

For Mercury assumed to be a rigid body reduced to three oblateness coefficients, 
the general expression for a torque coming from a disturbing point
mass $m$ at the vectorial position $r{\bf u}$ 
($r$ is the instantaneous distance) is written as follow~:
\begin{equation}
   {\bf N}_{2}=
   \left({3Gm\over a^{3}}\right)\left({a\over r}\right)^{3}
   {\bf u}\times (I)\,{\bf u} 
   \label{torque}
\end{equation}  
where $a$ is the mean distance between the two bodies.
Such a torque due to the Sun is the dominant one acting on the rotational
motion of Mercury. The solar torque is indeed responsible of the 
global dynamical behavior of the rotation of Mercury.
Due to Venus, the resulting torque is about of the order of $10^{5}$ times
smaller, that is quite negligible. 

Figure~\ref{SystSol} presents the rotational motion of Mercury including 
only the solar torque in the rotational equations and taking into account 
simultaneously the whole {\it N}-body problem for the Sun and the 
planets (the planetary interactions inducing indirect effects on the 
rotation of Mercury). In this Figure, the Euler angles $\psi,\theta,\varphi$
related to the 3-1-3 angular sequence describe the evolution of the body-fixed 
axes $Oxyz$ with respect to the axes of the local reference frame $OXYZ$. 
Let us recall the definition used for these angles~: $\psi$ is the precession
angle of the polar axis $Oz$ around the reference axis $OZ$, $\theta$ is the
nutation angle representing the inclination of $Oz$ with respect to $OZ$, and 
$\varphi$ is the rotation around $Oz$ and conventionally understood as the rotation
of the greatest energy (it is generally called the proper rotation). The axis of
inertia around which is applied the proper rotation is called the axis of figure
and defines the North pole of the rotation (Bois 1992). Let us remark that in
the Figure~\ref{SystSol} (in other Figures involving $\varphi$ as well) plotted
over 3000 years, we have removed the mean rotation of $58.646$ days in the
$\varphi$ angle in order to better distinguish the librations. We may then clearly
identify the first proper frequency of $15.847$ years (to be compared to
the analytical determination, namely 15.830 years given in Section \textit{2}). 

The $\psi$ angle expresses the nodal precession of the equatorial plane of Mercury
with respect to the ecliptic plane. It splits up in a periodic term with a period
$\Psi = 1066$ years and a secular one $\Pi$ = 278 898 years. $\Psi$ is the second
proper frequency of the Mercury's spin-orbit coupling. It can be analytically
approximated by the following formula (used in the Earth's case by
Goldstein $1981$)~:
\begin{equation}\label{prec}
       2\pi/\left(\frac{3}{2} \frac{n^{2}}{\omega_{z}}
       \frac{C-\overline{A}}{C} \cos{\theta}\right)
   \end{equation}
This analytical expression is suitable for an axis symmetric body. It is not the
case of Mercury. However, assuming Mercury as a symmetric top rotating
about its smallest axis of inertia, with an average equatorial moment of inertia
$\overline{A}=\frac{A+B}{2}$, one finds for this period 1300 years 
(by using the
values given in Table~\ref{paratot}). The difference between the two values 
permits to appreciate the departure of Mercury from a symmetric body. 
In the other hand, the dynamical behavior of $\psi$ coming from SONYR is due
to the direct effect of the solar torque by the way of the true dynamical 
figure of Mercury. In the 2-body problem, Sun-Mercury, $\Psi = 1066$ 
years and $\Pi = 0$ years. Related to the planetary interactions, $\Pi$ = 
278\,898 years expresses the departure from the 2-body problem.

\begin{figure}[!h]
      \begin{center}
        \hspace{0cm}
        \includegraphics[width=6cm]{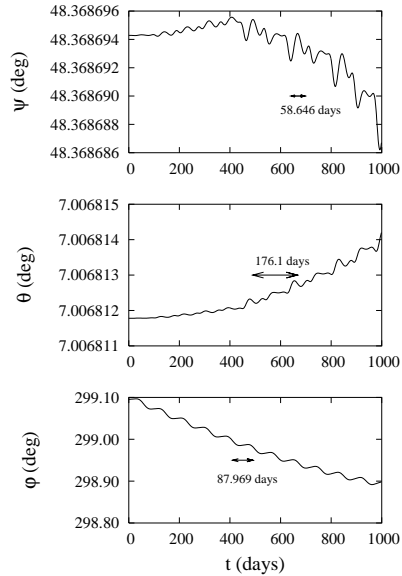}
	\vspace{-0.8cm}
        \caption{The rotational motion of Mercury expressed in the  
        ecliptic reference frame $OXYZ$ by using the 3-1-3 Eulerian 
        sequence $(\psi,\theta,\varphi)$, and plotted over 500 days. 
        Degrees are on the vertical axes and days on the horizontal axis.
        The periods of 58.646 and 87.969 days in the $\psi$ and $\varphi$ angles
        respectively, express precisely the 3:2 spin-orbit resonance 
        rate. The period $\tilde{P}$=176.1 days appears clearly 
        in these plots.}
        \label{stand500}
     \end{center}
\end{figure}

\begin{figure*}[!ht]
      \begin{center}
        \hspace{0cm}
        \includegraphics[width=10cm,angle=-90]{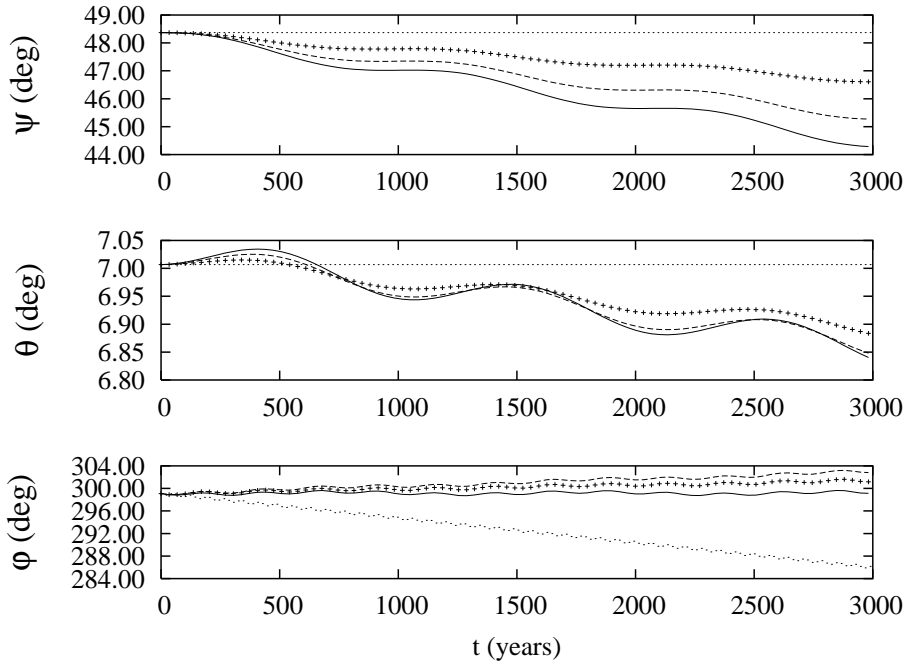}
        \vspace{-1.0cm}
        \caption{Interactions of the planets on the Hermean rotational 
        motion by the way of the spin-orbit couplings over 3000 years.
        Degrees are on the vertical axes and years on the horizontal axis 
        for the three panels.
        In the black line case, the problem is reduced to the Sun and 
        Mercury. In the broken line case, the interactions between the 
        Sun, Mercury and Venus are taken into account. With the dots
        line the later case includes Jupiter in addition. The whole 
        planetary interactions are integrated in the cross line case 
        (except for Pluto). The dots line case defines our called standard
        case sufficient for preserving the 3:2 resonance ratio.}
        \label{standard}
     \end{center}
\end{figure*}

The dynamical behaviors of $\psi$ and $\Omega$ (the ascending node of the orbit)
are quite superimposed as shown in Figure~\ref{standardCassini} (top panel).
As it is mentioned by B\'eletski (1986), a second synchronism is generally involved
in a Cassini state. However, in the Mercury's case, it is not a periodic synchronism
as it is the case for the Moon. Using SONYR, we make easily in evidence the 
periodic 18.6 year synchronism in the lunar spin-orbit resonance while it is about a
secular behavior in the Hermean spin-orbit resonance. The 3:2 resonance of Mercury 
introduces a mechanism of {\it spin-orbit secular resonance} qualitatively analogous
to the orbital secular resonances combined with mean motion resonances (as for
instance in the 2:1 case, see Bois \etal 2003). As it is shown in Figure~\ref{standardCassini}
(bottom panel), the spin-orbit secular resonance variable $\xi = \psi - \Omega$ does
not present any secular term. As a consequence, $\psi$ and $\Omega$ on average 
precess at the same rate equal to $\Pi$, confirming then the mechanism of spin-orbit 
secular resonance. We find that $\xi$ librates with the particular frequency
of 1066 years.

In addition, in order to give a detailed inspection of the short periods 
involved in the rotational motion of Mercury, Figure~\ref{stand500} 
presents the solution plotted over 500 days. The $P_{\varphi}$ rotation 
period of 58.646 days appears in the $\psi$ and $\theta$ angles. 
Whereas the mean rotation of 58.646 days is removed in the $\varphi$
angle (as in Figure~\ref{SystSol}), the signature of the $P_{\lambda}$
orbital period of 87.969 days is clearly visible (this angle is called libration
in longitude of 88 days in literature). A third period appears in the $\psi$ 
and $\theta$ angles, namely 176.1 days. This one results from the \32 
spin-orbit resonance ($\widetilde{P} = 2 P_{\lambda}=3 P_{\varphi}$).

Figure~\ref{standard} presents the planetary interactions acting 
on the rotational motion of Mercury by the way of its spin-orbit 
couplings (i.e. indirect effects of the planets on the Mercury's 
rotation). In the black line case the problem is reduced to the Sun and Mercury. 
In this 2-body problem, the orbital plane does not precess as it is 
clear in the $\psi$ and $\theta$ angles without secular terms. The secular
variations rise up from the departure of the 2-body problem (as it is 
visible with the broken, dots, and cross line cases in Figure~\ref{standard}). 
In the broken line case, the interactions between the Sun, Mercury and Venus
are taken into account. With the dots line, the later case includes Jupiter 
in addition. The whole planetary interactions are integrated in the cross line case 
(except for Pluto). We show that Venus is the planet which induces
the greatest secular term. After Venus, the role of Jupiter is dominant,
and this 4-body problem (Sun, Mercury, Venus, and Jupiter) defines our "standard" 
case used in our Section \textit{4.3} for the analysis of the Hermean librations.
The rate of secular variations in the Mercury's rotation between all 
planetary interactions (cross lines) and our standard case (dots lines) 
is $11.8$ as/cy (as~: arcseconds) in the $\theta$ nutation angle and $1.9$ amin/cy 
(amin~: arcminutes) in the 
$\psi$ precession angle. These values should be used as corrective 
terms in analytical theories of the rotational motion of Mercury.
Let us emphasize that the spin-orbit motion of Mercury coming from 
our standard case is sufficient for preserving the \32 resonance 
ratio between the two modes of motion.

\begin{figure}[!htb]
      \begin{center}
        \hspace{0cm}
        \includegraphics[width=6cm,angle=-90]{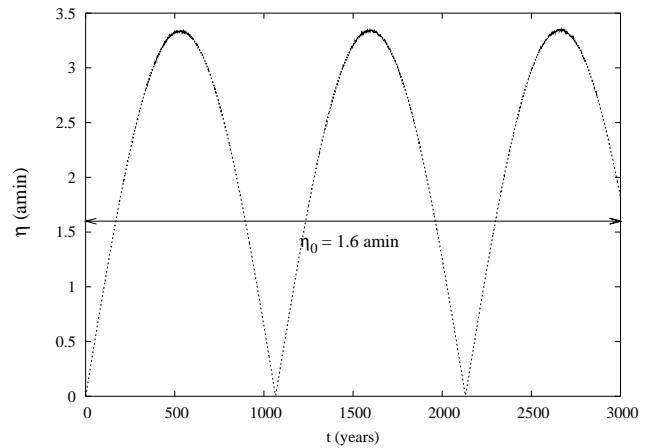}
        \caption{Dynamical behavior of the $\eta$ instantaneous obliquity 
        of Mercury plotted over 3000 years. Arcminutes are on the 
        vertical axis and years on the horizontal axis. Over this time of integration, 
        the behavior is simply described by a period of 1066 years and  
        3.2 amin of amplitude. The $\eta_{0}$ mean obliquity is 1.6 amin.}
        \label{obl_1}
      \end{center}
\end{figure}

Starting with the initial conditions defined in Section {\it 3.4} 
(where in particular the initial obliquity of Mercury is equal to 
zero), the SONYR model permits obtaining the dynamical behavior 
of the Hermean obliquity by its simultaneous spin-orbit integration.
The variables $i, \theta, \Omega, \psi$ from SONYR substituted
in the following relation~:
\begin{equation}
    \cos \eta = \cos i \cos\theta  + \sin i \sin\theta \cos(\Omega -\psi)
    \label{eq:obl}
\end{equation}
produce the instantaneous obliquity $\eta$ plotted over 3000 years
as presented in Figure~\ref{obl_1}. Such a behavior for $\eta$ gives 
easily a mean obliquity $\eta_{0}=1.6$ amin. We show at present that 
this mean value is quite consistent with the Cassini state of Mercury. 
Let be the following equation established by Peale (1988) and reformulated
by Wu \etal (1997) and coming from the Cassini laws~: 
\begin{equation}
      \frac{MR^{2}}{C} = \frac{\mu}{n} \frac{ \sin{(i +
      \eta_{0})}}{\sin{\eta_{0}} [ (1 + \cos{\eta_{0}}) G_{201}C_{22} -
      \cos{\eta_{0}}G_{210}C_{20}]}
     \label{eq:cassini}
\end{equation}
where $\mu$ is the precessional angular velocity of the Hermean orbit,
namely ${2 \pi}/{\Pi}$ $years^{-1}$, while 
    $$G_{201} = \frac{7}{2}e - \frac{123}{16} e^{3} \hbox{\ \ \ and \ \ }
    G_{210}= (1-e^{2})^{-\frac{3}{2}}$$ 
are eccentricity functions defined by Kaula (1966). $i$ is the inclination
of the orbital plane of Mercury relative to a reference system precessing 
with the orbit. This inclination varies between from $5^{\circ}$ to $10^{\circ}$
while the eccentricity varies from 0.11 to 0.24, over $10^{6}$ years (Peale 1988).
As a consequence, $\eta_{0}$ obtained with (\ref{eq:cassini}) belongs to [1.33, 2.65]
amin. Let us note that a conventional value of $7$ amin is often given in literature.
Such a value, outside the interval of possible values, is very probably incorrect as
already claimed by Wu \etal (1997). 


\subsection{Principal figure librations}

\subsubsection{Signature of the \cmr2 coefficient on the rotational 
motion of Mercury}

Let us consider at present the disturbing torques acting on the rotational 
motion of Mercury and as a consequence inducing direct librations. 
This section focuses on the librations related to the dynamical figure
of the planet. Such librations are called {\it principal figure librations}
(Bois 1995).
We assume the Sun reduced to a point mass while the gravity
field of Mercury is expanded in spherical harmonics up to the degree 2. 
We express the solar torque acting on the figure of Mercury according 
to the equation (\ref{torque}). 

The first coefficients of the Hermean gravity field have been determined
with the Mariner $10$ probe (Anderson \etal 1987). We use these values
for $C_{20}=-J_{2}$ and $C_{22}$ given in Table~\ref{paratot}.
In order to complete the Hermean tensor of inertia (coefficients $A, B, C$), 
the \cmr2 principal moment of inertia is required (see 
equation (\ref{ABC})). Its value is related to the internal density distribution 
of the planet according to the polar axis of Mercury (rotation of greatest energy
about the smallest principal axis of inertia). For an homogeneous planet, such
a normalized value is equal to 0.4. We use a nominal value of 0.34 
(Table~\ref{paratot}) used by Milani \etal (2001) and coming from an internal 
structure model of Mercury including three layers (crust, mantle and core).

\begin{figure}[!htb]
      \begin{center}
        \hspace{0cm}
        \includegraphics[width=6cm,angle=-90]{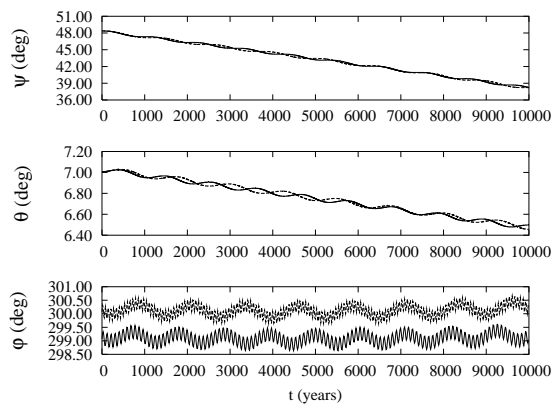}
        \vspace{-0.7cm}
        \caption{Rotational behavior of Mercury for two different 
        values of its greatest principal moment of inertia. Degrees are 
        on the vertical axes and years on the horizontal axes. 
        The computation with \cmr2=0.4 is plotted through the dashed
        lines; this value corresponds to an homogeneous planet.
        The broad lines are obtained with \cmr2=0.34; this value 
        derives from an internal structure model of Mercury including
        three layers. On the bottom panel ($\varphi$ angle), the 
        dashed lines are shifted away 1 degree in order to distinguish
        them from the broad lines.}
       \label{battement1}
       \end{center}
\end{figure}
\begin{figure}[!htb]
      \begin{center}
        \hspace{0cm}
        \includegraphics[width=6cm,angle=-90]{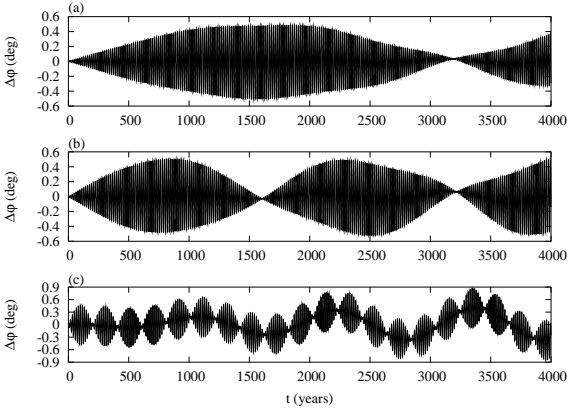}
        \vspace{-0.7cm}
        \caption{The signature of the libration of Mercury when the
        principal moment of inertia along the axis of figure varies from
        1 $\%$ in the top panel (a), to 2$\%$ in the mean panel and 15 
        $\%$ in the bottom panel (c). Degrees are on the vertical 
        axes and years on the horizontal axes. }
        \label{battement2}
       \end{center}
\end{figure}
\begin{figure*}[!]
      \begin{center}
        \hspace{0cm}
        \includegraphics[width=10cm,angle=-90]{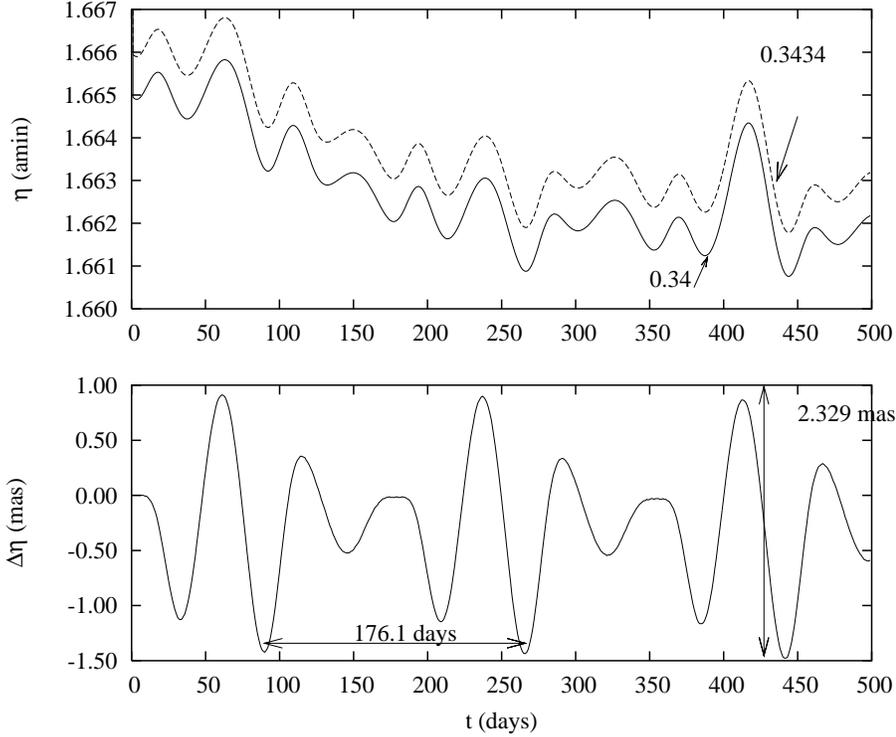}
        \caption{Signature of the \cmr2 coefficient on the obliquity 
        of Mercury for two different values shifted from $1\%$. 
        Arcminutes are on the vertical axis of the top panel while 
        milliarcseconds are on the vertical axis of the bottom 
        panel; days are on the horizontal axis for both panels. The maximal 
        amplitude of the signature of \cmr2 on the obliquity is 
        2.39 mas. On the top panel, the dashed lines are shifted away 
        0.001 amin in order to distinguish them from the broad lines.}
        \label{obl}
     \end{center}
\end{figure*}

Figure~\ref{battement1} presents the rotational behavior of Mercury 
computed over $10\,000$ years in our standard case with two different values
of its greatest principal moment of inertia~: (i) the computation with
\cmr2=0.4 is plotted with the dashed lines (homogeneous planet); 
(ii) the broad lines are obtained with \cmr2=0.34 (three layer model of Mercury). 
On the bottom panel ($\varphi$ angle), the dashed lines are shifted away
1 degree in order to distinguish them from the broad lines. 
This Figure~\ref{battement1} (especially the $\theta$ angle) shows how the \cmr2
coefficient value and the constant of precession $\Psi$ (2$^{nd}$ proper frequency) are related. 
With \cmr2=0.4, $\Psi= 1254.01$ years while with \cmr2=0.34, $\Psi=1066$ years.
Besides, \cmr2 and $\Phi$ (1$^{st}$ proper frequency) are also linked and for evaluating this 
relation, Figure~\ref{battement2} shows the variations $\Delta \varphi$
obtained by differentiation~: (i) on the top panel $\Delta C=1\%
$, (ii) on the middle panel $\Delta C=2\%
$, (iii) on the bottom panel $\Delta C=15\%
$ ($0.34 + 15\%
(0.34)=0.4$). The beats are signatures related to the variations in $\Phi$.

\subsubsection{Signature of the \cmr2 coefficient on the obliquity 
and on the libration angle in longitude}

One of the main objectives of the BepiColombo and MESSENGER missions
is to measure the rotation state of Mercury, up to an accuracy allowing
to constrain the size and physical state of the planet's core (Milani \etal 
2001, Solomon \etal 2001). Consequently, the two missions have to
determine the four following parameters~: $C_{20}$, $C_{22}$, $\eta$,
and $\varphi$ that are sufficient to determine the size and state of the
Mercury's core (see Peale 1988, 1997). Combining $C_{20}$, $C_{22}$,
and $\eta$, one obtains the \cmr2 coefficient while with $C_{22}$ and
$\varphi$, one obtains $C_{m}/MR^{2}$ (i.e. the \cmr2 coefficient for the
mantle). The validity condition of the first combination is that the dynamical
behavior of the core has to follow the one of the mantle over a period of time
at least the one of $\Pi$ (assertion 1). The validity condition of the second
combination is that the dynamical behavior of the core has not to be coupled
to the one of the mantle over a period of 88 days ($P_{\lambda}$) 
(assertion 2). These two conditions linked together imply some constraints
on the nature of the core-mantle interface (Peale 1997). In order to reach 
such an objective, the BepiColombo mission has to obtain a value on the 
\cmr2 coefficient with an accuracy of 0.003, i.e. $1\%
$ 
and therefore foresees measuring the libration angle and the obliquity with
an accuracy of 3.2 and 3.7 as respectively (Milani \etal 2001). 
\begin{figure*}[!]
      \begin{center}
        \hspace{0cm}
        \includegraphics[width=10cm,angle=-90]{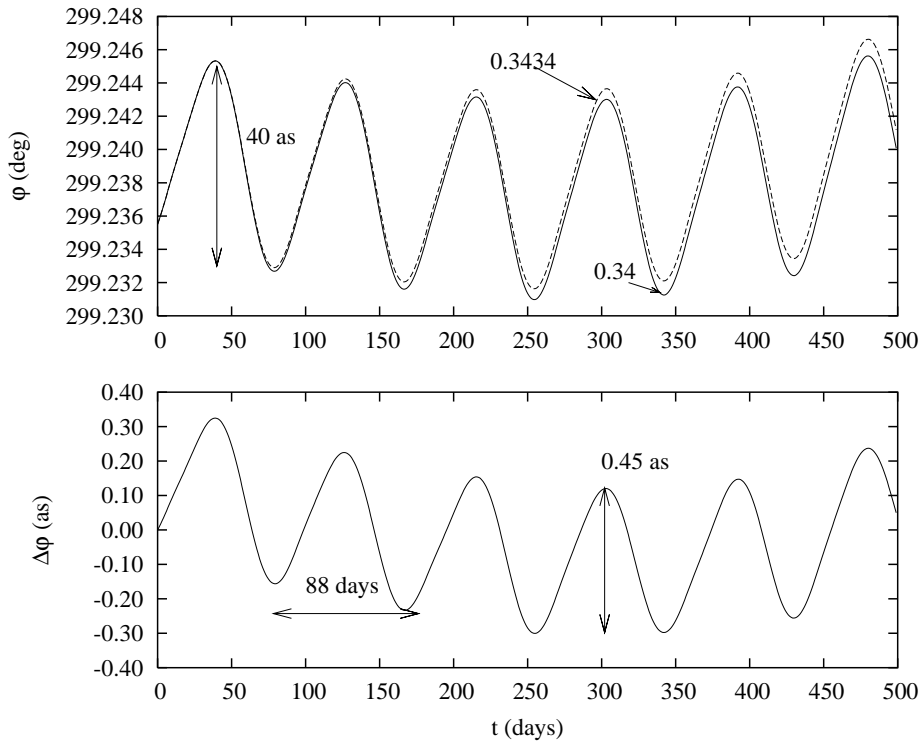}
        \caption{Signature of the \cmr2 coefficient on the Hermean libration angle
         in longitude for two different values shifted from $1\%$. 
        Degrees are on the vertical axis of the top panel and 
        arcseconds are on the vertical axis of the bottom 
        panel; days are on the horizontal axis for both panels. The maximal
        amplitude of the signature of \cmr2 on the libration angle in longitude is 0.45 as.}
        \label{lib}
     \end{center}
\end{figure*}

Our SONYR model gives (i) the true relation between the three 
parameters (\cmr2, $\eta$, $\varphi$), and (ii) the upper bounds 
of the impact of $C_{m}/MR^{2}$ on the $\varphi$ angle.
Figure~\ref{obl} presents the impact of the \cmr2 coefficient
on the instantaneous obliquity $\eta$. In these plots (Figures~\ref{obl}
and~\ref{lib}), the spin-orbit motion of Mercury is
integrated within the whole solar system with an initial obliquity of 
1.6 amin, which is the mean obliquity of Mercury evaluated in 
Section \textit{4.1}. The top panel of Figure~\ref{obl} expresses
the dynamical evolution of $\eta$ computed over 500 days with
\cmr2=0.34 (black lines) and \cmr2=0.3434 (dashed lines).
Dashed lines are shifted from 0.001 amin in order to distinguish
the two different kinds of lines. Figure~\ref{obl} shows also how the
instantaneous obliquity of Mercury differs from its $1.6$ amin
nominal value. The bottom panel shows by differentiation the signature
of the $1\%
$ variation of \cmr2 on $\eta$. The maximal amplitude crest to crest is
of the order of 2.3 mas within the characteristic period of 176.1 days 
related to the \32 ratio of the Mercury's spin-orbit resonance.

Figure~\ref{lib} presents the signature of the \cmr2 coefficient
on the $\varphi$ libration angle in longitude. The top panel expresses
the behavior of $\varphi$ computed over 500 days with
\cmr2=0.34 (black lines) and \cmr2=0.3434 (dashed lines).
One may compare this Figure~\ref{lib} to the Figure~\ref{plbg} resulting
from the usual analytical resolution of the Eulerian equation~\ref{bg}
(thanks to G. Giampieri,  private communication). Let us note that the 
angle $\gamma$ defined in Figure~\ref{plbg} is equal to the angle 
$\varphi$ plotted in Figure~\ref{lib}. The later only gives a simple 
double sine curve with an amplitude of 42 as while the solution of the 
SONYR model includes the couplings between the three rotational variables 
as well as the indirect couplings due to planetary interactions (we notice 
that in the two Figures~\ref{lib} and~\ref{plbg} the amplitude of libration
is of the order of 40 as). Let us note that the Figure~\ref{lib} corresponds
to the libration related to the \cmr2 coefficient of the planet without 
core-mantle couplings. Let us add that in Peale (1972), the amplitude
of $\varphi$ is related to the $C_{m}$ coefficient by assuming that
the assertion 2 quoted upper is true. On the contrary, our first results
on this topic make in evidence the existence of a faint coupling. 
This core-mantle coupling will be presented in a forthcoming paper. 
The bottom panel of Figure~\ref{lib} shows by differentiation
the signature of the $1\%
$ variation of \cmr2 on $\varphi$. The
maximal amplitude within the period of about 88 days
(i.e. the signature of $P_{\lambda}$) is of the order of $0.45$ as.

In conclusion, signatures of the indeterminacy of $1\%
$ in \cmr2 on the obliquity and on the libration in longitude are 2.4 mas 
and 0.45 as respectively. What is very faint (may be too much) with respect
to the expected accuracy forecasted in the BepiColombo mission.

\subsection{variing the obliquity}\label{sec:obliqui}

\begin{figure}[!htb]
      \begin{center}
        \hspace{0cm}
        \includegraphics[width=6cm,angle=-90]{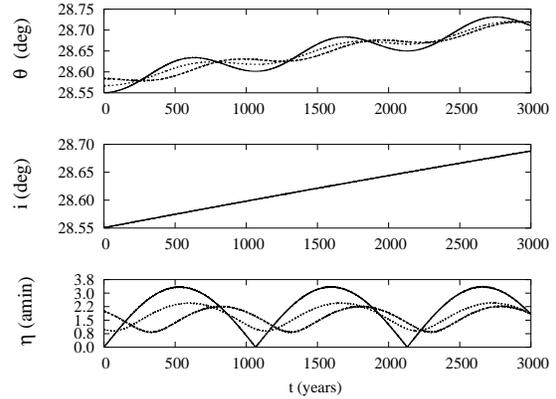}
	\vspace{-0.7cm}
        \caption{Impact of the initial obliquity on the nutation angle $\theta$ (top panel),
        on the orbital inclination $i$ (middle panel), and on the instantaneous obliquity 
        $\eta$ (bottom panel). Degrees are on the vertical axes of the top and middle
        panels while arcminutes are on the vertical axis of the bottom panel; years are
        on all horizontal axes. On each panel, three curves plotted over 3000 years 
        correspond to three different initial values of $\eta$~: 0 amin (black lines), 
        1 amin (dashed lines), and 2 amin (dot lines).}
        \label{varobl}
      \end{center}
\end{figure}

Because the initial obliquity value is unknown, we test in this last section
the impact of the indeterminacy of this value on the spin-orbit motion of
Mercury. The results are presented in Figure~\ref{varobl} plotted over 3000 
years; top panel: the effect on the nutation angle $\theta$, middle panel:
the effect on the orbital inclination $i$, and bottom panel: the effect on
the instantaneous obliquity $\eta$. On each panel, three curves are 
related to three different initial values of $\eta$, namely 0 amin (black lines), 
1 amin (dashed lines), and 2 amin (dot lines). 
In the bottom panel, the amplitudes of these librations are of the order
of 1.4 amin with a period of 1066 years. For any initial value of 
$\eta \in [0 , 3.2]$ amin, the mean value of $\eta$, let be $\eta_{0}$, is 
equal to 1.6 amin, which is in good agreement with the determination 
of $\eta_{0}$ in a previous section. We may claim that $\eta_{0}$=1.6 
amin.

For obtaining such a mean obliquity by measurements, let us underline 
that the theoretical behavior of $\eta$ points out to fit the observations 
by a sine function taking into account the long period $\Psi=1066$ years with
an amplitude of 1.6 amin.


\section{conclusion}

The 3:2 spin-orbit resonance between the rotational and orbital motions of
Mercury results from a functional dependance of the tidal friction adding 
to a non-zero eccentricity and a permanent asymmetry in the equatorial plane
of the planet. The upcoming space missions, MESSENGER and BepiColombo
with onboard instrumentation capable of measuring the rotational parameters 
stimulate the objective to reach an accurate theory of the rotational motion of
Mercury. 

Starting from our BJV relativistic model of solar system integration including
the coupled spin-orbit motion of the Moon, we have obtained a model 
generalizing the spin-orbit couplings to the terrestrial planets (Mercury, Venus,
Earth, and Mars). The updated model is called SONYR (acronym of Spin-Orbit N-BodY
Relativistic model). It permits to analyze and identify the different families of 
rotational librations. This work has been carried out for Mercury in the present 
paper. 

The spin-orbit motion of Mercury is characterized by two proper frequencies 
(namely $\Phi$ = 15.847 and $\Psi$ = 1066 years) and its 3:2 resonance presents a second 
synchronism which can be understood as a \textit{spin-orbit secular resonance}, 
($\Pi$ = 278\,898 years). A new determination of the mean obliquity has been proposed
in the paper. By using the SONYR model, we have found a mean obliquity of
1.6 amin. This value is consistent with the Cassini state of Mercury.  Besides, 
we have identified in the Hermean librations the impact of the uncertainty of the
greatest principal moment of inertia (\cmr2) on the obliquity and on the libration
in longitude (2.3 mas and 0.45 as respectively for an increase of 1$\%
$
on the \cmr2 value). These determinations prove to be suitable for providing
constraints on the internal structure of Mercury. The direct core-mantle
interactions will be presented in a forthcoming paper. 

 \begin{acknowledgements}
The authors thank Aleksei Pavlov for his help in the Poincar\'e 
cross-section computations and Jean Brillet for providing his 
efficient method of mean last squares useful for accurate 
determinations of periods in our data files.
 \end{acknowledgements}

\appendix 
\section{free rotation} \label{free}

Let us assumed Mercury isolated in space; in this sense, its rotation
is free and the Euler-Liouville equations for its rotation are written 
without right hand side, i.e. without any external disturbing torques.
If we add the assumption of a rigid Mercury, we are in the 
Euler-Poinsot motion case (whose solutions are the well-known Eulerian 
oscillations). Without explicitly integrating such equations, the assurance
of integrability in the Poincar\'e sense can be obtained by some 
theoretical simple considerations. Indeed, whatever being the triplet of
generalized coordinates used to describe the spatial attitude of a
solid body in a fixed frame, one knows that there exists four
independant integrals of motion~: the hamiltonian $H$, and the three
components $L_{X}, L_{Y}, L_{Z}$ of the angular momentum 
(in $OXYZ$). Four integrals of motion for three degrees of freedom,
the problem is then integrable and even over-integrable.
One does not lose the generality of the problem choosing for instance 
$A \leq B \leq C$. The choice  $A \leq B < C$ makes possible to write
the general solution of the system under a form involving the elliptical
functions of Jacobi (Landau \& Lifchitz 1969). By convention, let us
adopt that the resulting oscillations in space be called the \textit{Eulerian
oscillations}, expressing exclusively the oscillations of the non-perturbed
rotation of the rigid body. From this resolution, we obtain the Eulerian
frequencies~:
\begin{equation}
      \Omega_{f} = \sqrt{\frac{L^{2} -2AH}{C(C-A)}}
      \label{eq:fre1}
\end{equation}
and
\begin{equation}
      \sigma = \Omega_{f}  \sqrt{\alpha \beta}
      \label{eq:fre2}
\end{equation}
that give for Mercury the two periods $\Omega_{f}$ and $\sigma$
(where $L$ is the angular momentum in $OXYZ$, $H$ the energy).
$\alpha$ and $\beta$ are the dynamical coefficients of the body related to 
$A, B$ and $C$ by the following relations~:
\begin{equation}
      \alpha = \frac{C-B}{A}  \hbox{\ \ \ and \ \ } \beta = \frac{C-A}{B} 
      \label{coeffappla}
\end{equation}
With the previous initial conditions given in Table~\ref{paraci}
($\dot \psi= \dot \theta=0$), the free rotation of Mercury is reduced to 
an elementary rotation of the $\varphi$ angle only 
($\Omega_{f}=\omega_{z}$). Starting with the physical mean values
of $\dot \psi$, $\dot \theta$, and $\eta_{0}$ evaluated in the present paper,
$\Omega_{f}=58.646$ days and $\sigma=964.88$ years.
The period of $58.646$ days corresponds to the Hermean polar rotation
while the one of $964.88$ years means the global period of rotation of
Mercury in $R^{3}$, which is equivalent to the Earth's classical Euler period, 
namely 305 days (Lambeck 1980).  In the case of the Moon, this period is
equal to $148.129$ years.

\begin{figure}[!tbp]
      \begin{center}
        \hspace{0cm}
        \includegraphics[width=6cm,angle=-90]{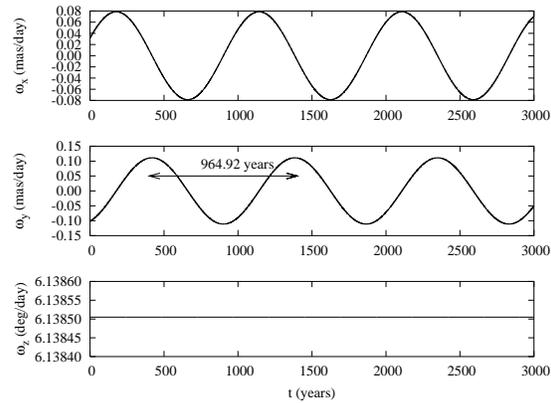}
	\vspace{-0.7cm}
        \caption{The free rotation of rigid Mercury in the components of 
        its instantaneous rotation vector expressed in the  
        body-fixed system $Oxyz$ and plotted over 3000 years. 
        Milliarseconds per day are on the vertical axis for $\psi$ 
        and $\theta$ angles, degrees per day for $\varphi$, and years on 
        the horizontal axes. The clearly visible period is the Euler period
	for Mercury~: namely 964.92 years. The mean value of $\omega_{z}$ 
	is 6.138 deg/day.}
        \label{bofree}
      \end{center}
\end{figure}

Using the SONYR model reduced to the free rotation of 
Mercury, we obtain the components of the instantaneous rotation vector 
${\bf \omega}$ in the body-fixed system $Oxyz$, as presented in 
Figure~\ref{bofree}. $\omega_{z}$ is well found constant while the Euler
period is equal to 964.92 years.

\end{document}